\documentclass[12pt]{article}

\usepackage{graphicx}

\begin{document}
\setlength{\baselineskip}{14pt}
\parindent=40 mm
\title{\Large\bf Axial and Vector correlator mixing\\
 in hot and dense hadronic matter}
\author{ G. Chanfray, J. Delorme, M. Ericson\thanks{ Also at
 Theory Division, CERN,
CH-1211 Geneva
 23, Switzerland} \\
{\small Institut de Physique Nucl\'{e}aire et IN2P3, CNRS,
Universit\'{e} Claude Bernard Lyon I,}
{\protect \vspace{-2mm}}
 \and {\small 43 Bd. du 11 Novembre,
F-69622 Villeurbanne Cedex, France} \\
 and\\
M. Rosa-Clot\\
{\small Dipartimento di Fisica, Universit\`a di Firenze e INFN,
 Firenze, Italia }\\}

\date{}

\maketitle

\begin{abstract}
We study the manifestations of chiral symmetry restoration which have a
significance for the parity mixing. Restricting to pions and nucleons we
establish a formalism for the expression of the vector correlator, which
displays the mixing of the axial correlator into the vector one and unifies the
cases of the heat bath and of the dense medium. We give examples of mixing
cross-sections. We also establish a link between the energy integrated mixing
cross-sections and the pion scalar density which governs the quenching factors
of coupling constants, such as the pion decay one, as well as the quark
condensate evolution.    
\end{abstract}

\centerline{PACS numbers: 11.30.Rd 11.40.-q 13.60.Fz 23.40.Bw 24.85.+p 25.20.Dc}
\vskip 4mm
\parindent = 30pt

\section{Introduction}

The consequences of chiral symmetry restoration in a dense or hot medium is one
of the fascinating problems raised in the sector of non perturbative QCD 
physics. At normal nuclear density for instance the amount of restoration is 
large (about 35\%)~\cite{DCL}. It is hardly conceivable that such a large effect
 does not show up in a way directly linked to the symmetry. 
These manifestations have been searched for either in the direction
 of mass changes or in parity related effects. 
For what concerns the second aspect, in the heat bath, Dey {\it et
al.}~\cite{DEY} (see also Ref.~\cite{STEEL}) showed that the thermal pions
induce a mixing of axial and vector correlators 
 accompanied by a depletion of the original correlator.
  At low temperature, this depletion is universal and equal to
   the mixing term. For a dense medium, Chanfray {\it et al.}~\cite{CDE}  
 established the existence of a certain analogy with the thermal case. One can
 consider that in a certain sense chiral symmetry restoration induces
  a mixing of a response of axial nature into the vector one, 
and {\it vice versa}, accompanied by a dropping 
of coupling constants such as that of the axial current. However the pions 
responsible for this mixing are those of the virtual nuclear pion cloud. 
The fact that they belong to the medium itself and not to an external
 reservoir introduces a major difference with the heat bath situation. In 
Ref.~\cite{CDE} the quenching factor was expressed  in terms of the scalar
 pion density,
which also partly governs chiral symmetry restoration. In this letter we
establish a formalism which treats on the same footing the heat bath and the 
dense case so as to reach a unified description when both effects are present, 
such as in heavy ions collisions. We illustrate some situations where the 
mixing pieces of the vector current act and we derive 
the mixing cross-sections that they
generate. We will show through the study of the Compton amplitude that, 
in the case of the nuclear medium, it is possible to
extract the quenching factor of the coupling constants from photoabsorption 
data. The importance of Compton scattering as a tool in relation to 
chiral symmetry was pointed out in Refs.~\cite{CE,ERK}.
 The existence of mixing cross-sections is linked to chiral symmetry
  restoration and can be viewed as one of its manifestations. 
Our conclusion is that such manifestations are 
numerous. It is only question to see familiar experimental data in a proper 
perspective. 

\section{Axial-Vector mixing in the nucleon and the nucleus}    
    
We start by rewriting the expression of the vector current as derived in 
Ref.~\cite{CDE} 
from a chiral Lagrangian in the Weinberg representation and in a world 
restricted to pions and nucleons:
\begin{eqnarray}
{\hbox{\boldmath${\mathcal V}$\unboldmath}}_{\mu} & = &
\frac{\displaystyle({\hbox{\boldmath$\phi
 \times $\unboldmath}}\partial_{\mu}{\hbox{\boldmath$\phi)
 $\unboldmath}}}{\displaystyle( 1 +
{\hbox{\boldmath$\phi $\unboldmath}}^2/4f_{\pi}^2)^2} \nonumber \\
& & +\frac{1}{2}\overline{\psi}\gamma_{\mu}
{\hbox{\boldmath$\tau $\unboldmath}}\psi
+\frac{1}{4f_{\pi}^2}\frac{\displaystyle\overline{\psi}\gamma_{\mu}
\big[{\hbox{\boldmath$(\tau\times\phi)\times\phi $\unboldmath}}\big]\psi}
{\displaystyle 1 +
{\hbox{\boldmath$\phi $\unboldmath}}^2/4f_{\pi}^2}
-\frac{g_A}{2f_{\pi}}\frac{\displaystyle\overline{\psi}\gamma_{\mu}\gamma_5
{\hbox{\boldmath$(\tau\times\phi) $\unboldmath}}\psi}
{\displaystyle 1 +
{\hbox{\boldmath$\phi $\unboldmath}}^2/4f_{\pi}^2}\; .
\label{vcur}
\end{eqnarray}

We remind which terms of expression~(\ref{vcur}) can be viewed as mixing terms.
 They are 
those which lead to a part of the axial current once a pion field in their
expression is taken care of by creation or annihilation of a thermal pion or a
nuclear virtual one. An example is the Kroll-Ruderman term ( last term of 
Exp.~(\ref{vcur})) which leads (within a normalization factor $1/f_\pi$)
 to the nucleonic
axial current. Another one is the photon coupling to the pion 
(term in \boldmath${\phi}$\unboldmath
 $\times\partial_{\mu}$\boldmath${\phi}$\unboldmath)
which leads to the creation or annihilation of a pion by the
axial current, once the pion field 
in it ({\it i.e.} the one without derivative) is removed. If this pion attaches 
to a nucleon then these two pieces together give the full axial current, with 
its induced pseudoscalar piece. It is important to stress that our definition
of mixing of the currents enlarges the notion of parity mixing. {\it Stricto
sensu} this is the mixing with states of opposite parity, as occurs for 
instance in the heat bath at low temperature, where the pions are soft.
For the mixing of the currents instead, the pions need not be soft, and 
indeed the nuclear pions are not (their momentum is a few hundred MeV/c). 

\begin{figure}
\begin{center}
\includegraphics[width=6 cm]{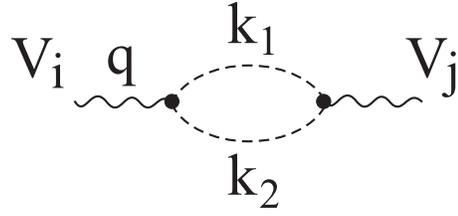}
\end{center}
\caption{\label{fig-fig0} Pionic contribution to the photon self-energy in the
vacuum.}
\end{figure}

\begin{figure}
\begin{center}
\includegraphics[width=14cm]{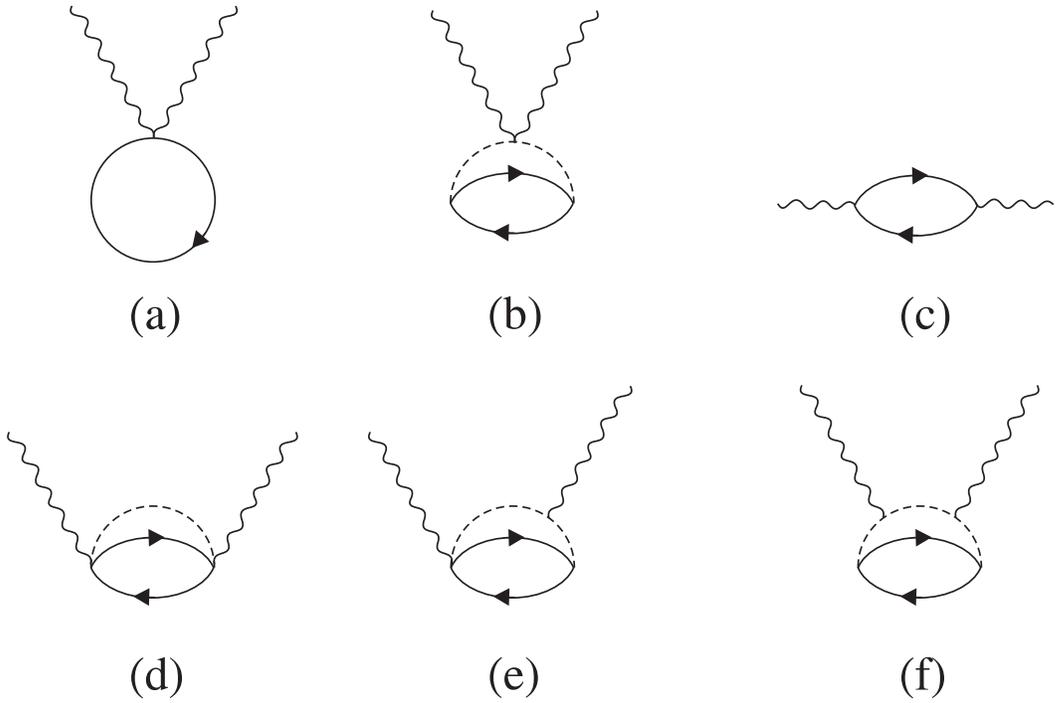}
\end{center}
\caption{\label{fig-fig1} The Compton amplitude on the nucleus to lowest order
in nucleon loops: seagull amplitudes (a-b), 
direct excitation of p-h by the vector current (c), mixing terms of the vector
current (d-f).}
\end{figure}

\begin{figure}
\begin{center}
\includegraphics[width=14cm]{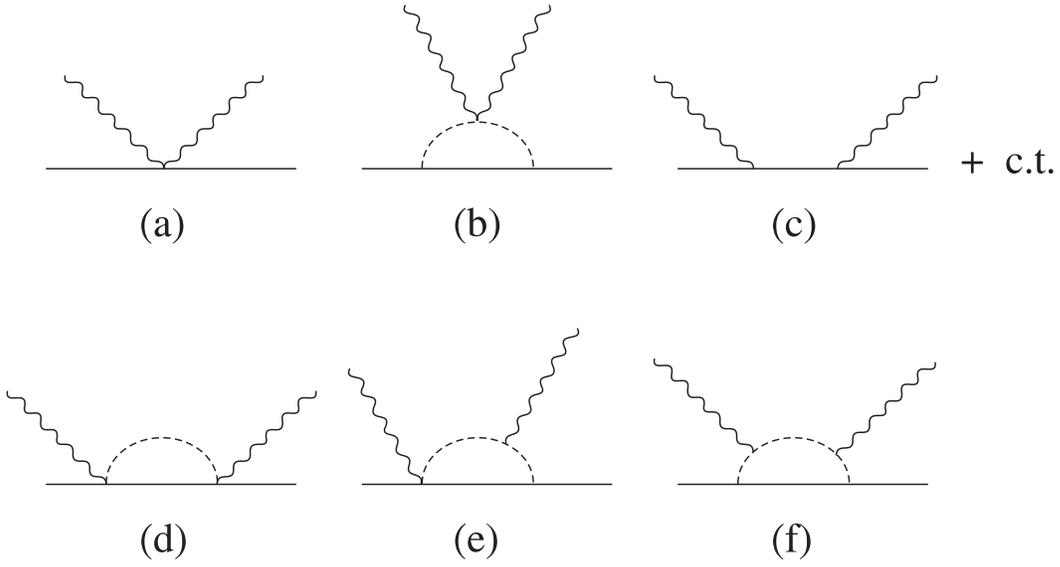}
\end{center}
\caption{\label{fig-fig2} The same as Fig.2 for the free nucleon.}
\end{figure}

We start by studying the propagation of photons in the nuclear medium. The case 
of the free proton will be discussed as well. It leads to an interesting 
possibility of an access to the pionic piece of the nucleon sigma commutator 
through photoabsorption data.
In the vacuum the photon self-energy contains the two-pions intermediate
states (Fig.~\ref{fig-fig0}). 
In the nuclear medium there is a direct excitation of particle-hole (p-h)
states by the current (graph 2c). In addition   
the pions are modified by the dressing by p-h states. It 
is also necessary to attach the photon to the vertices in order to satisfy
gauge invariance. In the one-nucleon 
loop approximation the corresponding graphs are
those of Fig.~\ref{fig-fig1}d-f. The graphs 2c-f which can be cut and have an imaginary
part represent the time ordered amplitude $T(\omega)$. Besides these terms the 
photon self-energy also contains seagull pieces: those where the two photons 
attach in a point-like fashion either to the nucleons or to the pions (graphs
2a-b). The piece 2a is given by the corresponding Thomson amplitude,
 $-e^2/M$ per proton with
$e^2=1/137$ and $M$ the proton mass. For the pions, which are virtual in 
the nucleus, the seagull amplitude can be expressed~\cite{ERK,NE},
 to lowest order in the pion fields, as the expectation value
of the squared pion field operator for charged pions: $ - \frac{2}{3} e^2 
\int d\vec{x}\, \langle A\vert{\hbox{\boldmath$\phi$\unboldmath}}(\vec{x})^2
\vert A\rangle$
, which is linked to the pion scalar density. Here and in the following $\vert
A\rangle $ stands for a nuclear state and $\vert N\rangle $ will be used for a
proton one. For a single proton the corresponding graphs for the forward 
photon-proton amplitude are shown in Fig.~\ref{fig-fig2}.
 We start by discussing this case. Summing the graphs of Fig.~\ref{fig-fig2}
 we decompose the $\gamma N$ forward spin independent amplitude as:
\begin{equation}
f_N(\omega) = - e^2/M + S_\pi + T(\omega)\ .
\label{fn}
\end{equation}
In the low energy limit, this amplitude should take the Thomson value, $-e^2/M$. 
Thus the dressing of the nucleon by the pion does not affect
 the low energy value. This constraint imposes the following relation 
between the time ordered amplitude and the seagull term:
 $ S_\pi + T(0) = 0$.
In the absence of form factors this relation is indeed satisfied by the 
calculated amplitudes from the graphs 3b-f, which is no surprise because
the introduction of these graphs is necessary to satisfy gauge invariance, 
i.e., the low energy theorem. For instance in the static approximation, we 
have:\vfill\break
\begin{eqnarray}
  S_\pi&=  &- e^2 \int d\vec{x}\,
\langle N\vert\big[{\hbox{\boldmath$\phi$\unboldmath}}(\vec{x})^2
    - \phi_3(\vec{x})^2\big]\vert N\rangle = - \frac{2}{3} e^2 \int d\vec{x}\,
 \langle N\vert{\hbox{\boldmath$\phi$\unboldmath}}(\vec{x})^2\vert N\rangle
  \nonumber \\
  & = &-\frac{e^2g^2}{4\pi^2 M^2}\int_0^\infty dq\,\frac{q^4}{(q^2+m_\pi^2)^2} 
        \nonumber  \\
T(0) & = & \frac{e^2 g^2}{4\pi^2 M^2}\int_0^\infty dq\,
\big[\frac{q^2}{q^2+m_\pi^2}-\frac{4}{3}\frac{q^4}{(q^2+m_\pi^2)^2}
+\frac{4}{3}\frac {q^6}{(q^2+m_\pi^2)^3}\big] \ ,      
\label{fstat}       
\end{eqnarray}        
 where the three terms in the last integrand correspond to the three
  graphs 3d-f and $g$ is the $\pi NN$ coupling constant. The
 positive energy nucleon Born term (graph 3c) does not contribute at
 $\omega = 0$. In the sum $S_\pi + T(0)$
  the divergent pieces cancel and the remaining integrals give an overall
  vanishing result.      
  
More generally, writing a once subtracted dispersion relation for the proton
Compton amplitude, we get:
\begin{equation}
 Re f_N(\omega) = Re f_N(0) + \frac{2\omega^2}{\pi}\int_0^\infty
  d\omega'\, \frac{Imf_N(\omega')}{\omega'(\omega'^2-\omega^2)} =
 -\frac{e^2}{M} + \frac{\omega^2}{2\pi^2}
\int_0^\infty d\omega'\, \frac{\sigma_\gamma(\omega')}{\omega'^2-\omega^2} \ ,
\label{disrel}
\end{equation}
where $\sigma_{\gamma}$ is the photoabsorption cross-section 
which corresponds to the graphs 3d-f.
It arises from the Kroll-Ruderman and photoelectric terms which originate 
from the mixing pieces of the vector current.
The corresponding cross-section thus 
represents the mixing of the axial correlator into the vector one.
In the high energy limit, $\omega \to \infty$ (defined as high compared to
the excitation energies for the relevant degrees of freedom, see
Ref.~\cite{ERK} for a detailed discussion), $f_N$ reduces to its
seagull parts:
\begin{equation}
  f_N(\infty) = - \frac{e^2}{M} - \frac{2}{3}e^2 \int d\vec{x}\,
  \langle N\vert{\hbox{\boldmath$\phi$\unboldmath}}(\vec{x})^2\vert N\rangle \ .
 \label{fnty}   
 \end{equation}
 Inserting~(\ref{fnty}) into~(\ref{disrel}) we obtain the relation:
 \begin{equation}
 \int d\vec{x}\,
 \langle N\vert{\hbox{\boldmath$\phi$\unboldmath}}(\vec{x})^2 \vert N\rangle
     = \frac{3}{2e^2}\,\frac{1}{2\pi^2}\int_0^\infty d\omega'\, \sigma_\gamma(\omega') \ .
 \label{phi2}
 \end{equation} 
  It is then possible
  to evaluate the expectation value of the squared 
pion field from the part of the photoabsorption cross-section arising from
the Kroll-Ruderman and pionic Born terms. The direct excitation of the
 $\Delta$ resonance
 is not included in this description. 
It has to be subtracted from the cross-section. However the $\Delta$ has to be 
included as an intermediate state in the graphs of Fig.~\ref{fig-fig2} (except
\ref{fig-fig2}c).
 Indeed it is known 
that a proper description of the pion cloud needs the $\Delta$-pion intermediate 
state. This is no difficulty. The seagull (pionic) 
term remains formally the same: $ - \frac{2}{3} e^2 \int d\vec{x}\,
 \langle N \vert{\hbox{\boldmath$\phi$\unboldmath}}(\vec{x})^2\vert N\rangle$.
  As 
for the cross-section it has to include the pion photoproduction accompanied 
by $\Delta$ excitation of the nucleon. This cross-section has been 
measured~\cite{PID} and 
we use it for our experimental input. It is strongly suppressed above 1-1.2 
GeV, where other channels, not incorporated in our description, open. This 
provides the scale at which a description in terms of one pion, eventually 
accompanied by $\Delta$ excitation ceases to be valid. We use the same cut off 
 for the non-$\Delta$ part. We thus evaluate a total pion number (defined as the 
volume integral of the scalar pion density) 
 $N_\pi = m_\pi\int d\vec{x}\,
 \langle N\vert {\hbox{\boldmath$\phi$\unboldmath}}(\vec{x})^2\vert N\rangle$                        
of about 0.35. 
Another idea of this magnitude is provided by transforming this number into the 
pionic piece of the nucleon sigma commutator, using the relation: 
$\Sigma_N^{(\pi)} = \frac{1}{2}\, m_\pi^2 
\int
d\vec{x}\,\langle N\vert{\hbox{\boldmath${\phi}$\unboldmath}}(\vec{x})^2\vert
N\rangle \approx 25$ MeV (out of this, 9 MeV come from the $\Delta$).
This value well agrees with theoretical estimates~\cite{JCT,BMG},
 showing the consistency of our overall concept of the meaning of mixing cross
 sections. As the sigma commutator provides a measure of the 
amount of chiral symmetry restoration brought in by nucleon, according 
to the relation:
\begin{equation}
 \Sigma_N = 2m_q\int d\vec{x}\, 
\big[\langle N\vert \overline{q}q(\vec{x})\vert N\rangle
-\langle 0\vert \overline{q}q(\vec{x})\vert 0\rangle\big] \ ,
\label{Sigma}
\end{equation}
 the equation~(\ref{phi2})
amounts to a link between the amount of 
chiral symmetry restoration of pionic origin and the mixing cross-sections.

\begin{figure}
\begin{center}
\includegraphics[width=14cm]{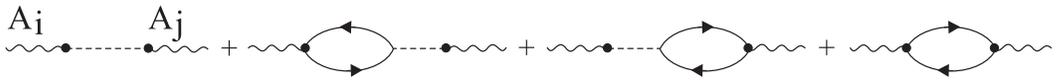}
\end{center}
\caption{\label{fig-fig3} The axial correlator $A_{ij}$.}
\end{figure}

We now turn to the nuclear case. In the one-nucleon loop 
approximation the graphs 
which enter in the photon self-energy are those of Fig.~\ref{fig-fig1}.
 Those which correspond to mixing terms of the current 
are those of Fig.~\ref{fig-fig1}d-f where the photon attaches 
to the $NN\pi$ vertex or to a pion in flight. Their contribution can be
 concisely 
introduced in the vector correlator, as explained below. We restrict ourselves
 to the pure space components, the other one can be obtained through current
 conservation. Let us first consider the axial correlator $A_{ij} (i,j = 1,2,3)$
 as represented in Fig.~\ref{fig-fig3}. Since our illustration of the mixing is
 made on that of the axial correlator into the vector one, we have ignored the
 mixing terms of the axial current. The axial current then couples only to the
 nucleon and the pion. In
 the non relativistic limit and for a spin saturated system the correlator 
 $A_{ij}$ thus truncated writes:
\begin{eqnarray}
\frac{1}{f_\pi^2}A_{ij}(k) &= &  k_i k_j D(k) + 2k_i k_j\Pi_0(k)D(k) 
+ \hat{k}_i\hat{k}_j\Pi_L(k) + (\delta_{ij}-\hat{k}_i\hat{k}_j)\Pi_T(k)
   \nonumber   \\ 
      &= & k_i k_j\big(1 + \Pi_0(k)\big)^2 D(k) +\hat{k}_i\hat{k}_j\Pi_0(k)
      + (\delta_{ij}-\hat{k}_i\hat{k}_j)\Pi_T(k) \ .
      \label{acorr}
\end{eqnarray}
As for the vector correlator we skip the third term in the 
expression~(\ref{vcur}) of the vector current which is a mixing term but would
lead to a part of the axial current that we have ignored ({\it i.e.} the
Weinberg-Tomozawa term). We also neglect the
pion fields in the denominators. The correlator then follows from the graph of
Fig.~\ref{fig-fig1}c-f and writes (see also Ref.~\cite{CS}):
\begin{eqnarray}
 V_{ij}(q)& = & {\Pi_V}_{ij}(q) + i\int \frac{d^4k_1}{(2\pi)^4}\,
  \big[\big(1+\Pi_0(k_1)\big)k_{1i} -  \big(1+\Pi_0(k_2)\big)
  k_{2i}\big]   \nonumber    \\
  &  & \times D(k_1)D(k_2)\big[\big(1+\Pi_0(k_1)\big)k_{1j} - 
  \big(1+\Pi_0(k_2)\big) k_{2j}\big] \nonumber  \\
  &  & + i\int\frac{d^4k_1}{(2\pi)^4}\, \big[ \hat{k}_{1i}\hat{k}_{1j}\Pi_0(k_1)D(k_2) +
   (\delta_{ij}-\hat{k}_{1i}\hat{k}_{1j})\Pi_T(k_1)D(k_2) \nonumber \\
  &  & + \hat{k}_{2i}\hat{k}_{2j}\Pi_0(k_2)D(k_1) +
   (\delta_{ij}-\hat{k}_{2i}\hat{k}_{2j})\Pi_T(k_2)D(k_1)\big]  \ ,   
\label{vcorr}
\end{eqnarray}

where $k_1$ and $k_2$ are the two pions four-momenta, (their sum is the incident
momentum $q = k_1+k_2 $) and ${\Pi_V}_{ij}$ is the RPA response to 
(longitudinal) vector excitations (Fig.~\ref{fig-fig1}c).
 In the two expressions~(\ref{acorr},\ref{vcorr})
 $D_0$ is the free pion propagator, $D$ is the fully
dressed one with p-h insertions and $\Pi_L$ and $\Pi_T$ are the RPA responses to
the isospin and spin excitations, respectively for the spin longitudinal
and spin transverse ones. The index 0 for the quantities $\Pi$ corresponds to
the irreducible pion self-energy (including effects of short range
correlations which can be parametrized by the usual $g'$ parameter)  
 and we have: $D = D_0 + \vec{k}^2 D_0\Pi_0 D$. 
With these expressions for the correlators, we do not restrict to the
one-nucleon loop appoximation but the full RPA chains are included. This means
that in the graphs 2 and 4, the pion is considered 
as a quasi-particle with a full
dressing of p-h insertions. In the vector correlator, we separate out the part
 with only one nucleon-hole propagator (Fig.~\ref{fig-fig1}c) and its RPA chain
 which has nothing to do with mixing.   
The remainder can be expressed in terms of the axial correlator, 
in a form which displays the mixing effect:
\begin{eqnarray}
V_{ij}(q) - {\Pi_V}_{ij}(q) & = & i\int \frac{d^4k_1}{(2\pi)^4}\,
 \big[ \frac{1}{f_\pi^2}\big(A_{ij}(k_1)D(k_2)
  + A_{ij}(k_2)D(k_1)\big) \nonumber \\
  - &\big(1{\hbox{\hskip-5.mm}} &+\Pi_0(k_1)\big)\big(1+\Pi_0(k_2)\big)(k_{1i}k_{2j}+k_{2i}k_{1j})
  D(k_1)D(k_2)\big]    \ .
\label{mixing}
\end{eqnarray}
Indeed, if one of the pions 1 or 2 is a thermal one, or a pion from the
nuclear pion cloud (in this case the pion propagator $D$ has to be dressed
by at least one p-h insertion), one is left with the axial correlator,
taken at the momentum of the other pion. On the right hand side of
Eq.~(\ref{mixing}) there is an extra term which does
not reduce to the product of the axial correlator with the pion
propagator. Its existence is due to the interaction of the photon with the
pion (term in \boldmath${\phi}$\unboldmath$\times\partial_{\mu}$\boldmath
${\phi}$\unboldmath). 
The derivative can act on either of
the two pions. The action on the thermal or the nuclear one leads to this
extra piece. Indeed, as well known, in the photoelectric
 part of pion production on a
nucleon, the momentum dependence is not that of the exchanged pion
$k$, but it involves the sum of the two outgoing and exchanged pions, 
{\it i.e.}, $2k-q$, while the pseudoscalar piece of the axial current contains
 only the exchanged pion momentum $k$. However the extra term of the vector
correlator, which bars the simple factorization, does not alter the basic
mixing concept. Indeed, when one pion is taken as a thermal or nuclear one,
the remainder is still of axial nature. In the large incident momentum limit,
 when $q$ is larger than the momentum
distribution of the thermal or nuclear pion, the other pion practically
carries the external momentum which is larger and the terms in $k_1 k_2$ can be
 ignored. In this case the factorization holds.

The expressions~(\ref{acorr}-\ref{mixing}) unify 
the dense and thermal
cases. In the latter one, the polarization propagators $\Pi$ have to be ignored
and the integration over $k_1$ has to be understood as a sum over Matsubara  
frequencies. The formalism applies equally well in the mixed case where both
 density and temperature effects are present,
  such as in heavy ion collisions.

 Coming back to the nuclear case the spin independent Compton amplitude $f_A$ 
 ( ${\hbox{\boldmath$\epsilon\cdot\epsilon'$\unboldmath}}f_A =
  - \epsilon_i\epsilon'_j V_{ij}$),
 is also constrained by the low energy theorem: $f_A(\omega = 0) = -
 (Ze)^2/MA$. We make the decomposition into 
 the seagull terms $S$ (graphs 1a-b) and the time ordered part $T$ (graphs~
 1c-f), where $S= S_N + S_{pion}$ with
\begin{equation}
S_N  =  - \frac{Ze^2}{M} \hbox{\phantom{aaaa} and\phantom{aaaa}} 
S_{pion} =  - \frac{2}{3} e^2 \int d\vec{x}\,
 \langle A\vert{\hbox{\boldmath$\phi$\unboldmath}}(\vec{x})^2\vert A\rangle  \ .
 \label{Snuc}
 \end{equation}
In the high energy limit, only the seagull terms survive. A once subtracted
dispersion relation then provides the pionic seagull term:
\begin{equation}
- S_{pion} = S_N - f_A(0) + \frac{1}{2\pi^2}\int_0^\infty d\omega'\,
\sigma_{\gamma A}(\omega') \ ,
\label{Spion}
\end{equation} 
from which we get \footnote {Similar ideas have been put forward by
Gerasimov~\protect\cite{GER}}   
, per nucleon:
\begin{equation}
 \frac{1}{A} \int d\vec{x}\,
 \langle A\vert{\hbox{\boldmath$\phi$\unboldmath}}(\vec{x})^2\vert A\rangle =
 \frac{3}{2e^2}\big[-\frac{ZN}{MA^2} + \frac{1}{2\pi^2}\int_0^\infty d\omega'\,
\frac{\sigma_{\gamma A}(\omega')}{A}\big] \ .
\label{SpionA}
\end{equation} 

As mentioned previously, the direct $\Delta$ excitation should be removed from 
the photoabsorption cross-section. In the energy 
integrated cross-section we assume its
 contribution to be the same as in the nucleon, as suggested in Ref.~\cite{CC}. 
 Using the same cut-off ($\approx 1$ GeV) as for the free proton 
we find, as in Ref.~\cite{ERK}, a slightly increased value as compared to the
free nucleon corresponding to a (scalar) pion number of 0.38 ({\it vs} 0.35 in
the free case) at the normal 
nuclear density $\rho_0$. A certain amount of increase is indeed expected from  
the spin-isospin correlations in the nucleus. We remind that the expectation
value defined by relation~(\ref{SpionA}) governs, in the
nuclear medium, the quenching factor $r$ of coupling constants such as 
the pion decay one~\cite{CDE}, for which we obtain  $r(\rho_0) \approx 0.86$.           

\section{Conclusion}

 We conclude with some remarks about the manifestations of chiral symmetry
  restoration. We have seen that the cross-sections where the photon
  attaches to a pion or to a pion-nucleon state through the Kroll-Ruderman term
  can be viewed as mixing cross-sections. The interpretation of certain 
  cross-sections as mixing ones is reinforced by the relations that we have
   established in Eqs.~(\ref{phi2},\ref{SpionA})
 between them and the pion scalar density. The latter
governs the condensate evolution as well as the quenching factor of certain
coupling constants. A natural question is whether these manifestions of chiral
symmetry restoration, mixing and depletion effects, can be identified in
experiments. In the nucleus the nucleon which is the source of
 the pion field responsible for the mixing, has to be ejected. This 
automatically leads to final states in the continuum, such as 2p-2h excitations,
 with the possibility for one particle to be a $\Delta$.
  The mixing cross-sections thus have a broad energy distribution. Associated
  with the mixing is a quenching effect, which  
 can apply to discrete states, or a group of them. 
 An example is the Gamow-Teller sum 
 rule where the states of the giant Gamow-Teller (GT) resonance are excited. The 
 depopulation of these states for the benefit of the continuum of the mixing 
 cross-section is thus a manifestation of chiral symmetry restoration. This 
 implies that the states of the giant GT resonance do not exhaust the 
 axial strength. The vector correlator which is introduced by the mixing term 
 of the axial current provides some additional broad strength which
 compensates, partly or totally, the depletion of strength of the giant GT 
 states linked to chiral symmetry. Both signal the mixing and chiral
  symmetry restoration (however other nuclear processes such as core
  polarization can have a similar effect, for a review see Ref.~\cite{TOW}). It is possible 
 that the missing strength of the longitudinal response in (e,e') scattering 
 follows the same pattern. It was indeed suggested~\cite{ME} that the
 quasi-elastic longitudinal strength is depopulated in favour of the
 nucleon-hole $\Delta$-hole continuum. Another case is that of 
 the heavy ion collisions where the inclusion of p-h excitations, 
 {\it i.e.} mainly of the graphs of Fig.~\ref{fig-fig1}d, is responsible for a large 
 part of the excess at low mass electron 
 pairs~\cite{CS,CRW}. This also represents a mixing cross-section 
 {\it i.e.} a manifestation of chiral symmetry restoration. 
 The consistency of the description implies the corresponding depletion 
 of the $\rho$ meson excitation by the vector current due to pion loops 
 to be simultaneously taken into account. This quenching should not be only the
 thermal one considered for instance in Ref.~\cite{SLK}, but it has to
incorporate also the presence of the baryonic background. We are pursuing
our work in this direction.

To summarize our findings we can conclude that 
in the nuclear medium, the mixing of axial and vector correlators, enlarged to 
the nuclear case and defined in Eq.~(\ref{mixing}),
 has a signature because the 
nuclear pions have a broad energy and momentum distribution. We have given a 
formalism where the analogy between the dense case and the heat bath is 
apparent. In the case where the original correlator concerns a group of narrow 
states, chiral symmetry restoration implies their depopulation and the 
appearance of the mixed correlator which has a broad energy distribution.

\bigskip
We thank Prof. S.B. Gerasimov for useful discussions. We are also grateful to
Dr. D. Davesne and J. Marteau for stimulating interaction.

\newpage

\section*{Figure captions}

Figure 1~: Pionic contribution to the photon self-energy in the vacuum.
\vskip 4mm
\noindent Figure 2~: The Compton amplitude on the nucleus to lowest order
in nucleon loops: seagull amplitudes (a-b), 
direct excitation of p-h by the vector current (c), mixing terms of the vector
current (d-f).
\vskip 4mm
\noindent Figure 3~: The same as Fig.2 for the free nucleon.
\vskip 4mm
\noindent Figure 4~: The axial correlator $A_{ij}$.
\newpage

\end{document}